\documentstyle[pra,aps,epsfig]{revtex}
\input epsf

\newcommand{\news}{\setcounter{equation}{0}}
\newcommand{\be}{\begin{equation}}
\newcommand{\ee}{\end{equation}}
\newcommand{\bea}{\begin{eqnarray}}
\newcommand{\eea}{\end{eqnarray}}
\newcommand{\bean}{\begin{eqnarray*}}
\newcommand{\eean}{\end{eqnarray*}}
\font\upright=cmu10 scaled\magstep1
\font\sans=cmss12
\newcommand{\ssf}{\sans}
\newcommand{\stroke}{\vrule height8pt width0.4pt depth-0.1pt}
\newcommand{\Z}{\hbox{\upright\rlap{\ssf Z}\kern 2.7pt {\ssf Z}}}

\newcommand{\C}{{\rlap{\rlap{C}\kern 3.8pt\stroke}\phantom{C}}}
\newcommand{\R}{\hbox{\upright\rlap{I}\kern 1.7pt R}}
\newcommand{\CP}{\C{\upright\rlap{I}\kern 1.5pt P}}
\newcommand{\half}{\frac{1}{2}}
\newcommand{\mt}{\rlap{\ssf T}\kern 3.0pt{\ssf T}}

\newcommand{\identity}{{\upright\rlap{1}\kern 2.0pt 1}}

\begin{document}
\pagestyle{plain}

\preprint{DAMTP-97-55, UKC/IMS/96-70}
\title{SU(N) Monopoles and Platonic Symmetry\footnote{DAMTP-97-55,
    UKC/IMS/96-70, hep-th/9708006}} 
\author{Conor J. Houghton\thanks{Electronic address:
    C.J.Houghton@damtp.cam.ac.uk}}
\address{DAMTP, Silver Street, Cambridge, CB3 9EW, U.K.}
\author{Paul M. Sutcliffe\thanks{Electronic address:
    P.M.Sutcliffe@ukc.ac.uk}}
\address{Institute of Mathematics,
University of Kent at Canterbury, Canterbury CT2 7NZ, U.K.}
\date{June 1997} 
\maketitle

\maketitle
\begin{abstract}
We discuss the ADHMN construction for $SU(N)$ monopoles
and show that a particular simplification arises in studying
charge $N-1$ monopoles with minimal symmetry breaking. 
Using this we construct families of tetrahedrally
symmetric $SU(4)$ and $SU(5)$ monopoles. In the moduli space
approximation, the $SU(4)$ one-parameter family describes a novel
dynamics where the monopoles never separate, but rather, a tetrahedron
deforms to its dual. We find a two-parameter family of $SU(5)$ tetrahedral
monopoles and compute some geodesics in this submanifold numerically.
The dynamics is rich, with the monopoles scattering either
once or twice through octahedrally symmetric configurations.
\end{abstract}

\newpage
\section{Introduction}
\indent\ Bogomolny-Prasad-Sommerfield monopoles are static solitons occurring
in certain (3+1)-dimensional gauge field theories. They have attracted
interest continually since they were discovered over two decades ago.
The simplest BPS monopoles are $SU(2)$ monopoles, which have an
associated integer referred to as the topological charge. Monopoles
with topological charge $k$ are called $k$-monopoles, and their total
energy is proportional to $k$. For any given $k$ there are
$k$-monopole solutions which resemble $k$ well-separated 1-monopoles.

It is interesting, but often difficult, to examine monopoles associated with
gauge groups larger than $SU(2)$. Such monopoles have
features not found in the $SU(2)$ case, for example, the existence of
spherically symmetric multi-monopoles \cite{BW}.

There is a powerful approach to monopoles; the ADHMN construction. To
perform this construction a nonlinear differential equation, called
the Nahm equation, must be solved and its solution, the Nahm data,
used to define a linear ordinary differential equation.  This linear
equation, which we shall refer to as the ADHMN construction equation,
must then be solved. Its solutions yield the fields via an integration
procedure.  Recently Platonic symmetries have been exploited to
construct Nahm data and these symmetric Nahm data have been used to
examine some particular examples of $SU(2)$ monopoles. In this paper
we discuss two cases where the same Platonic Nahm data, slightly
modified, can be used to study monopoles associated with larger gauge
groups.

Section \ref{MNd} is an introduction to $SU(N)$ Nahm data and
monopoles. We show that for minimal symmetry
breaking the Nahm data for some multi-monopoles is 
simpler than the corresponding $SU(2)$ Nahm data.
This allows us, in a simple way,
to modify known $SU(2)$ Nahm data so that it is $SU(N)$ Nahm data. It
is possible to use $SU(2)$ Nahm data to produce $SU(2)$ monopoles
embedded in an $SU(N)$ theory. Such embedded monopoles behave like
$SU(2)$ monopoles. 
The modification we consider is more
radical than a simple embedding and the corresponding monopoles behave
quite unlike the way $SU(2)$ monopoles do. 

In Section \ref{Smwts} we use charge three Nahm data
with tetrahedral symmetry to construct a one-parameter family of
$SU(4)$ monopoles with tetrahedral symmetry. 
The dynamics of slow moving monopoles is approximated by geodesic
motion in the moduli space of solutions \cite{M}. Since the one-parameter
family of solutions described in Section \ref{Smwts} is the fixed
point set of the action of the tetrahedral group, it must be a
geodesic. Thus the one-parameter family described in Section
\ref{Smwts} is an example of a pathological scattering process in which the
monopole never separates into distinct objects.

In Section \ref{Smwts2} we use charge four Nahm data with tetrahedral
symmetry to construct a two-parameter family of $SU(5)$ monopoles with
tetrahedral symmetry. This two-parameter family is totally geodesic in
the whole moduli space. Under the assumption that the transformation
between the metric on the space of Nahm data and  the metric on the moduli
space of monopoles is an isometry, we 
undertake a numerical study of the low energy dynamics of $SU(5)$
tetrahedral monopoles. We find an exotic dynamics involving both single
and double scatterings through configurations with octahedral symmetry.

\section{Monopoles and Nahm data}\label{MNd}
\ \indent BPS $SU(N)$ monopoles are topological solitons in an $SU(N)$ Yang-Mills-Higgs gauge
theory with no Higgs self-coupling. They are finite energy solutions to the Bogomolny
equation
\be
D_i\Phi=-\half\epsilon_{ijk}F_{jk}
\label{bog}
\ee
where $D_i=\frac{\partial}
{\partial x_i}+[A_i,$ \ is the covariant derivative with $A_i$ an 
{\sl su(N)}-valued gauge potential and $F_{jk}$ the gauge field.
 $\Phi$ is an {\sl su(N)}-valued scalar field, called the Higgs
field. Non-trivial
asymptotic conditions are imposed on the Higgs field, which are responsible
for the existence of topological soliton solutions to the theory.
It is required
that, as $r=\vert {\bf x}\vert$ approaches infinity, $\Phi$ takes values in the
gauge orbit of the matrix
\be M=i\,\mbox{diag}\,(\mu_1,\mu_2,\ldots,\mu_N).\ee
By convention it is assumed that $\mu_1\le\mu_2\le\ldots\le\mu_N$. Since
$\Phi$ is traceless $\mu_1+\mu_2+\ldots+\mu_N=0$. This $M$ is the
vacuum expectation value for $\Phi$ and the
symmetry group of $M$ under gauge transformation is called the
residual, or unbroken,
symmetry group. Thus, for example, if all the $\mu_p$ are
different the residual symmetry group is the maximal torus
$U(1)^{N-1}$. This is known as maximal, or generic, symmetry
breaking. 
The soliton solutions are associated with $N-1$
integers; this is because the boundary condition on $\Phi$
implies a map, $\Phi_{\infty}$, from the large sphere at infinity into the
quotient group
\be\Phi_{\infty}:S^2\rightarrow
\mbox{orbit}_{SU(N)}M \ =SU(N)/U(1)^{N-1}\ee
and
\be
\pi_2\left(SU(N)/U(1)^{N-1}\right)=\pi_1(U(1)^{N-1})=\Z^{N-1}.
\ee
In contrast, this paper concerns the minimal symmetry breaking case,
 in which all but one of the $\mu_p$ are identical, so the
residual symmetry group is $U(N-1)$. It
is convenient to choose $\mu_1=-(N-1)$ and $\mu_p=1$ for $p=2\ldots
N$. Since
\be
\pi_2\left(SU(N)/U(N-1)\right)=\Z
\ee 
there is only one topological integer associated with
solutions. Nonetheless, a given solution has $N-1$ integers associated
with it, these arise in the following way; careful analysis of the
boundary conditions indicates that there is a choice of gauge such
that the Higgs field for large $r$, in a given direction, is given by
\be\Phi(r)=i\,\mbox{diag}\,(\mu_1,\mu_2,\ldots,\mu_N)-\frac{i}{2r}\,\mbox{diag}\,(k_1,k_2,\ldots,k_N)+O(r^{-2}).\ee
In the maximal symmetry breaking case the
topological charges are given by 
\be m_p=\sum_{q=1}^pk_q.\ee
In the case of minimal symmetry breaking only the first of these
numbers, $m_1$, is a topological charge. Nonetheless, the remaining
$m_p$ constitute an integer characterization of a solution. This
characterization is gauge invariant up to reordering of the
integers $k_p$. The $m_p$ are known as magnetic weights, with the matrix
 $\mbox{diag}\,(k_1,k_2,\ldots,k_N)$ often called the
charge matrix.
 
There are some obvious ways of embedding $su(2)$ in $su(N)$, for example,
\be \left(\begin{array}{cc}\alpha&\beta\\-\bar{\beta}&-\alpha\end{array}\right)\hookrightarrow\left(\begin{array}{ccccc}\ddots&&&&\\ &\alpha&\ldots&\beta& \\
&\vdots&\ddots&\vdots&\\ &-\bar{\beta}&\ldots&-\alpha& \\&&&&\ddots\end{array}\right).\label{emb}\ee
Important $SU(N)$ monopoles can be produced by
embedding the $SU(2)$ 1-monopole fields, which are known $su(2)$-valued fields, in
$su(N)$. Some care must be taken in producing these embedded monopoles to ensure
that the asymptotic behaviour is correct; the $SU(2)$ monopole may
need to be scaled and it may be necessary to add a
constant diagonal field beyond the plain embedding described by
(\ref{emb}), details may be found in \cite{W,Wa}. Obviously there is an
embedding of the form (\ref{emb}) for each choice of two columns in
the target matrix. The embedded 1-monopoles have a single $k_p=1$ and
another $k_p=-1$, the rest are zero. The choice of columns for
the embedding dictates which two $k_p$ are non-zero.
Recall that in the case of minimal symmetry breaking the choice of order of
the $k_p$ is a gauge choice.
In fact, in the case of minimal symmetry breaking, the embedded
1-monopole is unique up to position and gauge
transformation. Solutions with $k_1=k$ have $k$ times the energy of this
basic solution and so it is reasonable to call these $k_p=k$ monopoles
$k$-monopoles. There are of course different types of such $k$-monopoles
corresponding to different magnetic weights. 

Consider $SU(3)$ monopoles with minimal symmetry breaking. For $k=2$
there are two distinct types of monopoles, those
with $m_2=0$ and those with $m_2=1$. The $m_2=2$ case is equivalent to
the $m_2=0$ case; $m_2$ can be changed from 0 to 2 by reordering $k_2$
and $k_3$. If $m_2=0$ the monopoles are all
embeddings of $su(2)$ 2-monopoles and this case is not
interesting as an example of $su(3)$ 2-monopoles. The
$m_2=1$ case has been studied by Dancer \cite{D} and by Dancer and
Leese \cite{DL}, by considering Nahm data.     

There is an equivalence between Nahm data and BPS monopoles. In the case
of $SU(N)$ monopoles the Nahm data are triplets of anti-hermitian matrix functions
$(T_1,T_2,T_3)$ of $s$ over the intervals $(\mu_p,\mu_{p+1})$. The size of
the matrices depends on the corresponding values of $m_p$; the
matrices $(T_1,T_2,T_3)$ are $m_p\times m_p$ matrices in the interval
$(\mu_{p},\mu_{p+1})$. They are required to be non-singular in each
region and to satisfy the Nahm equation
\be
\frac{dT_i}{ds}=\half\epsilon_{ijk}[T_j,T_k]. \label{Neqn}
\ee   
There are complicated boundary conditions prescribed between the Nahm
data in abutting intervals, which are detailed in Nahm's original paper
\cite{Nahm}. We will follow Hurtubise and Murray's formulation of the
Nahm data boundary conditions for distinct $\mu_p$ \cite{HM} and then take the
limit of coincident $\mu_p$ to describe the minimal symmetry breaking case.

For ease of notation we shall describe the case where $m_{p-1}\ge m_p$
(ie. $k_p\le 0$ for $p>1$) since a similar result holds after a reordering
if this is not satisfied.

Monopoles are constructed from their corresponding Nahm data by first
solving a first order differential equation in which the Nahm data
appear as coefficients. This is called the ADHMN construction
equation. This choice affects the order of the $k_p$. Rather than describe it in full generality, it will be
described below in the particular form required. The matching and boundary conditions on Nahm
data are designed to ensure that the ADHMN construction equation has the
 correct number of solutions required to yield the correct type of monopole fields. 
Define the function
\be k(s)=\sum_{p=1}^Nk_p\,\theta(s-\mu_p)\ee
where $\theta(s)$ is the usual Heaviside function. In the interval
$(\mu_p,\mu_{p+1})$ $k(s)=m_p$. It is a rectilinear
skyline whose shape depends on the charge matrix of the corresponding monopole. 
If $k(s)$ near $\mu_p$ is
\begin{center}
\setlength{\unitlength}{0.012500in}%
\begingroup\makeatletter\ifx\SetFigFont\undefined
\def\x#1#2#3#4#5#6#7\relax{\def\x{#1#2#3#4#5#6}}%
\expandafter\x\fmtname xxxxxx\relax \def\y{splain}%
\ifx\x\y   
\gdef\SetFigFont#1#2#3{%
  \ifnum #1<17\tiny\else \ifnum #1<20\small\else
  \ifnum #1<24\normalsize\else \ifnum #1<29\large\else
  \ifnum #1<34\Large\else \ifnum #1<41\LARGE\else
     \huge\fi\fi\fi\fi\fi\fi
  \csname #3\endcsname}%
\else
\gdef\SetFigFont#1#2#3{\begingroup
  \count@#1\relax \ifnum 25<\count@\count@25\fi
  \def\x{\endgroup\@setsize\SetFigFont{#2pt}}%
  \expandafter\x
    \csname \romannumeral\the\count@ pt\expandafter\endcsname
    \csname @\romannumeral\the\count@ pt\endcsname
  \csname #3\endcsname}%
\fi
\fi\endgroup
\begin{picture}(360,121)(100,524)
\thinlines
\put(300,540){\line( 0,-1){  6}}
\put(330,600){\vector( 0, 1){  0}}
\put(330,600){\vector( 0,-1){ 55}}
\put(250,640){\vector( 0, 1){  0}}
\put(250,640){\vector( 0,-1){ 94}}
\multiput(403,605)(8.22222,0.00000){5}{\line( 1, 0){  4.111}}
\put(180,645){\line( 1, 0){120}}
\put(300,645){\line( 0,-1){ 40}}
\put(300,605){\line( 1, 0){100}}
\put(400,605){\line(-1, 0){  1}}
\put(100,540){\line( 1, 0){380}}
\multiput(179,645)(-8.66667,0.00000){5}{\line(-1, 0){  4.333}}
\put(314,624){\makebox(0,0)[lb]{\smash{\SetFigFont{12}{14.4}{rm}$-k_p$}}}
\put(310,645){\vector( 0, 1){  0}}
\put(310,645){\vector( 0,-1){ 36}}
\put(286,524){\makebox(0,0)[lb]{\smash{\SetFigFont{12}{14.4}{rm}$s=\mu_p$}}}
\put(254,590){\makebox(0,0)[lb]{\smash{\SetFigFont{12}{14.4}{rm}$m_{p-1}$}}}
\put(335,570){\makebox(0,0)[lb]{\smash{\SetFigFont{12}{14.4}{rm}$m_p$}}}
\put(445,602){\makebox(0,0)[lb]{\smash{\SetFigFont{12}{14.4}{rm}$k(s)$}}}
\end{picture}
\end{center}
then as $s$ approaches $\mu_p$ from below we require 
\be T_i(s)=\left(\begin{array}{cc}\frac{1}{z}R_i+O(1) &
O(z^{(|k_p|-1)/2}) \\O(z^{(|k_p|-1)/2})  & T_i^{\prime}+O(z)
\end{array}\right)
\label{frombelow}
\ee
where $z=s-\mu_p$ and where
\be
T_i(s)=T_i^{\prime}+O(z)\ee
as $s$ approaches $\mu_p$ from above.

It follows from the Nahm equation (\ref{Neqn}) that the $|k_p|\times |k_p|$ residue matrices $(R_1,R_2,R_3)$ in
(\ref{frombelow}) form a representation of $su(2)$. The boundary
conditions require that this representation is the unique irreducible
$|k_p|$-dimensional representation of $su(2)$. 

In summary, at the boundary between
two abutting intervals, if the Nahm matrices are $m_{p-1}\times m_{p-1}$ on the
left and $m_{p}\times m_{p}$ on the right  an
$m_p\times m_p$ block continues through the boundary and there is an
$(m_{p-1}-m_p)\times(m_{p-1}-m_p)$ block simple pole whose residues
form an irreducible representation of $su(2)$. The boundary conditions for $m_{p-1}=m_p$ are
given in, for example, \cite{HM}. While these boundary conditions are
involved, their function is simply one of limiting the number of solutions to
the ADHMN construction equation.

The 1-dimensional representations of $su(2)$ are trivial. Thus, if
$k_p=-1$ for all $p>1$, $k(s)$ is   
\begin{center}
\begin{picture}(341,100)(159,520)
\thinlines
\put(310,609){\makebox(0.1111,0.7778){\SetFigFont{5}{6}{rm}.}}
\multiput(305,600)(8.46154,0.00000){7}{\line( 1, 0){  4.231}}
\multiput(360,600)(0.00000,-8.00000){3}{\line( 0,-1){  4.000}}
\multiput(360,580)(7.27273,0.00000){6}{\line( 1, 0){  3.636}}
\put(400,580){\line( 1, 0){ 20}}
\put(420,580){\line( 0,-1){ 20}}
\put(420,560){\line( 1, 0){ 40}}
\put(460,560){\line( 0,-1){ 15}}
\put(460,545){\line( 0,-1){  5}}
\put(460,540){\line( 1, 0){ 40}}
\put(260,540){\line( 0,-1){  6}}
\put(200,540){\line( 0, 1){ 80}}
\put(200,620){\line( 0, 1){  0}}
\put(200,620){\line( 0, 1){  0}}
\put(200,620){\line( 1, 0){ 60}}
\put(260,620){\line( 0,-1){ 20}}
\put(260,600){\line( 1, 0){ 40}}
\put(300,600){\line( 0, 1){  0}}
\put(300,600){\line( 0, 1){  0}}
\put(190,615){\vector( 0, 1){  0}}
\put(190,615){\vector( 0,-1){ 68}}
\put(159,540){\line( 1, 0){301}}
\put(420,540){\line( 0,-1){  6}}
\put(173,574){\makebox(0,0)[lb]{\smash{\SetFigFont{12}{14.4}{rm}$k_1$}}}
\put(460,540){\line( 0,-1){  6}}
\put(200,540){\line( 0,-1){  6}}
\put(195,524){\makebox(0,0)[lb]{\smash{\SetFigFont{12}{14.4}{rm}$\mu_1$}}}
\put(255,524){\makebox(0,0)[lb]{\smash{\SetFigFont{12}{14.4}{rm}$\mu_2$}}}
\put(415,524){\makebox(0,0)[lb]{\smash{\SetFigFont{12}{14.4}{rm}$\mu_{N-1}$}}}
\put(455,524){\makebox(0,0)[lb]{\smash{\SetFigFont{12}{14.4}{rm}$\mu_{N}$}}}
\end{picture}
\end{center}
and the Nahm data has only one pole, it is at $s=\mu_1$. It seems
reasonable to suppose that this result holds in the limit of
coincident $\mu_p$. Thus, if we fix $\mu_1=-(N-1)$ and
$\mu_{p\not=1}=1$, we expect that the $N\times N$ Nahm data whose sole
pole is at $s=-(N-1)$, satisfying the Nahm equations and having
acceptable residues, are the Nahm data of $SU(N)$ monopoles with
minimal symmetry breaking. For this to be the case it is only required
that the ADHMN construction equation have the correct number of
solutions over the interval. This index calculation is easily
performed using the methods of \cite{HM}. The topological charge of
the corresponding monopole solution is, of necessity, $k_1=N-1$, since
the $k_p$ must add to zero. The magnetic weights are each one less
than the proceeding one. We say that the magnetic weights are
distinct. It has recently been proven by Nakajima that {\em all}
monopoles of this type can be constructed from the described Nahm
data\cite{Na}.

The Nahm data for $su(2)$ $k$-monopoles are $k\times k$ anti-hermitian matrix
solutions of the Nahm equation. They have poles at $s=-1$ and $s=1$. It
is obvious that this Nahm data,
\begin{center}
\begin{picture}(201,100)(159,520)
\put(180,580){\makebox(0,0)[lb]{\smash{\SetFigFont{12}{14.4}{rm}$k$}}}
\thinlines
\put(330,614){\vector( 0, 1){  0}}
\put(330,614){\vector( 0,-1){ 69}}
\put(333,580){\makebox(0,0)[lb]{\smash{\SetFigFont{12}{14.4}{rm}$k$}}}
\put(360,540){\line( 0, 1){  0}}
\put(190,614){\vector( 0, 1){  0}}
\put(190,614){\vector( 0,-1){ 69}}
\put(200,540){\line( 0,-1){  6}}
\put(320,540){\line( 0,-1){  5}}
\put(159,540){\line( 1, 0){201}}
\put(200,540){\line( 0, 1){ 80}}
\put(200,620){\line( 1, 0){120}}
\put(320,620){\line( 0,-1){ 80}}
\put(320,540){\line( 0, 1){ 35}}
\put(317,524){\makebox(0,0)[lb]{\smash{\SetFigFont{12}{14.4}{rm}1}}}
\put(190,524){\makebox(0,0)[lb]{\smash{\SetFigFont{12}{14.4}{rm}$-1$}}}
\put(365,540){\makebox(0,0)[lb]{\smash{\SetFigFont{12}{14.4}{rm},}}}
\end{picture}
\end{center}
 can be used to generate Nahm data for
$SU(k+1)$ $k$-monopoles with distinct weights
\begin{center}
\begin{picture}(201,100)(159,520)
\thinlines
\put(200,540){\line( 0, 1){ 80}}
\put(200,620){\line( 1, 0){120}}
\put(320,620){\line( 0,-1){ 45}}
\put(320,575){\line( 0, 1){  0}}
\multiput(320,570)(0.00000,-8.57143){4}{\line( 0,-1){  4.286}}
\put(315,580){\line( 1, 0){  5}}
\put(315,600){\line( 1, 0){  5}}
\put(360,540){\line( 0, 1){  0}}
\put(200,540){\line( 0,-1){  6}}
\put(180,575){\makebox(0,0)[lb]{\smash{\SetFigFont{12}{14.4}{rm}$k$}}}
\put(320,540){\line( 0,-1){  5}}
\put(159,540){\line( 1, 0){201}}
\put(190,614){\vector( 0, 1){  0}}
\put(190,614){\vector( 0,-1){ 69}}
\put(330,619){\vector( 0, 1){  0}}
\put(330,619){\vector( 0,-1){ 18}}
\put(330,599){\vector( 0, 1){  0}}
\put(330,599){\vector( 0,-1){ 18}}
\put(317,524){\makebox(0,0)[lb]{\smash{\SetFigFont{12}{14.4}{rm}1}}}
\put(190,524){\makebox(0,0)[lb]{\smash{\SetFigFont{12}{14.4}{rm}$-k$}}}
\put(365,540){\makebox(0,0)[lb]{\smash{\SetFigFont{12}{14.4}{rm}.}}}
\put(334,587){\makebox(0,0)[lb]{\smash{\SetFigFont{12}{14.4}{rm}1}}}
\put(334,607){\makebox(0,0)[lb]{\smash{\SetFigFont{12}{14.4}{rm}1}}}
\end{picture}
\end{center}
In examples where the charge $k$ $SU(2)$ data is known, the
$SU(k+1)$ data is generated by a translation and rescaling of $s$
so that a pole occurs at $s=\mu_1$ but the second pole is moved outside
the interval $s\in[\mu_1,\mu_N]$ ie. it is lost from the Nahm data.
The 2-monopole Nahm data is known exactly, and was used by Dancer in \cite{D} 
to construct $SU(3)$ monopoles. This is the simplest application of the
above procedure. Platonic
symmetry groups have previously been used to derive higher charge Nahm
data, and in this paper we discuss the corresponding $SU(k+1)$ monopoles.

\section{SU(4) monopoles with tetrahedral symmetry}\label{Smwts}
\news
In the previous Section we discussed the Nahm data for
$SU(k+1)$ monopoles with minimal symmetry breaking, charge
$k$ and distinct magnetic weights. For the remainder of the paper it
will be convenient to perform a translation $s\mapsto s-\mu_1$, so that
a pole in the Nahm data always occurs at $s=0$.
In this Section we describe
some aspects of the
the ADHMN construction, which calculates the monopole 
fields $(\Phi,A_i)$ from the Nahm data. We then go on to 
apply this construction to obtain a one-parameter family
of monopoles with $k=3$, which have tetrahedral symmetry.\\

Given
Nahm data $(T_1,T_2,T_3)$ for a $k$-monopole we must solve the 
ADHMN construction equation, for $s\in[0,k+1]$,
\be
({\identity}_{2k}\frac{d}{ds}+{\identity}_k\otimes x_j\sigma_j
+iT_j\otimes\sigma_j){\bf v}=0
\label{lin}
\ee
for the complex $2k$-vector ${\bf v}(s)$, where $\identity_k$ denotes
the $k\times k$ identity matrix, $\sigma_j$ are the Pauli matrices and
${\bf x}=(x_1,x_2,x_3)$ is the point in space at which the monopole
fields are to be calculated. Introducing the inner product
\be
\langle{\bf v}_1,{\bf v}_2\rangle =\int_0^{k+1} {\bf v}_1^\dagger{\bf v}_2\ ds
\label{ip}
\ee
then the solutions of (\ref{lin}) which we require are those which are
normalizable with respect to (\ref{ip}). It can be shown that the
space of normalizable solutions to (\ref{lin}) has (complex) dimension
$k+1$. If $\widehat {\bf v}_1,..,\widehat {\bf v}_{k+1}$ 
is an orthonormal basis
for this space then the $ij$th matrix element, $(\Phi)_{ij}$,
of the Higgs field is given by
\be
(\Phi)_{ij}=i
\langle(s-k)\widehat {\bf v}_i,\widehat {\bf v}_j\rangle 
\label{higgs}
\ee
A similar expression exists for the gauge potential, but the
energy density, ${\cal E}$, may be computed 
without calculating the
gauge potential by using the formula
\be
{\cal E}=\bigtriangleup \mbox{tr}(\Phi^2)
\label{lap}
\ee
where $\bigtriangleup$ denotes the Laplacian on \R$^3$.

For most of the examples considered in this paper the
Nahm data is sufficiently complicated that the matrix linear
differential equation (\ref{lin}) can not be solved 
analytically in
closed form. In these cases we use a numerical implementation
of the ADHMN construction which involves solving the 
ordinary differential equations using a standard fourth
order Runge-Kutta method. The numerical implementation is
similar to that introduced by the authors in \cite{HSa},
but is simplified by the fact that the Nahm data we
consider here has a pole at only one end of the $s$ interval.
This eliminates the need for the shooting part of the
numerical algorithm described in \cite{HSa}.

In references \cite{HMM,HSa,HSb} it is explained how 
Platonic symmetry (that is, tetrahedral, octahedral
or icosahedral symmetry) may be applied to Nahm data
for $SU(2)$ monopoles of charge $k$. We use these, as explained in the
previous Section, to easily obtain the solutions to 
Nahm's equation for $SU(k+1)$ monopoles.

From \cite{HSa} we have that the Nahm data for tetrahedrally
symmetric monopoles with $k=3$ has the form
\be
T_1=\left[\begin{array}{ccc}
0&0&0\\
0&0&-z\\
0&\bar z&0
\end{array}
\right]
\;\;
T_2=\left[\begin{array}{ccc}
0&0&-\bar z\\
0&0&0\\
z&0&0
\end{array}
\right]
\;\;
T_3=\left[\begin{array}{ccc}
0&z&0\\
-\bar z&0&0\\
0&0&0
\end{array}
\right]
\ee
where 
\be
z=\frac{\omega\wp^\prime(\omega s)}
{2\wp(\omega s)}+\frac{\sqrt{3}\omega}{\wp(\omega s)}
\mbox{ , }\;\;
\omega=e^{i\pi/6}\kappa
\ee
and 
$\wp$ is the Weierstrass function satisfying
\be
\wp^{\prime2}=4\wp^3-4
\ee
where $^\prime$ denotes differentiation with respect 
to the argument.

For $SU(2)$ monopoles the boundary condition requires
that the Nahm data has a simple pole at $s=2$ which
determines $\kappa$ to be
\be
\pm\kappa=\kappa_0=\frac{\Gamma(1/6)\Gamma(1/3)}
{12\sqrt{\pi}}.
\ee
For $SU(4)$ monopoles the corresponding boundary condition
is less restrictive, namely that we require the Nahm data
to have no poles for $s\in(0,4]$. Given the $SU(2)$ result
this implies that $\kappa$ must satisfy the condition
\be
-\kappa_0/2<\kappa<\kappa_0/2.
\ee
It is a simple task to show that the remaining 
requirements for Nahm data are satisfied, and hence
we have proved the existence of a one-parameter family
of $SU(4)$ monopoles with tetrahedral symmetry.
The one-parameter family is given by $\kappa$ in the 
above interval and the corresponding family of spectral 
curves is
\be
\eta^3+i
36\kappa^3
\zeta(\zeta^4-1)=0. \label{sca}
\ee

Note that $\kappa=0$ gives the spectral curve
$\eta^3=0$, which is the spectral curve of a 3-monopole
 with spherical symmetry. 
Such a spherically symmetric monopole was 
constructed several years
ago \cite{BW} by using a spherically symmetric ansatz
in the field equations. We shall now see how this 
solution arises in the ADHMN construction by explicitly
calculating the Higgs field in this case.

Taking the limit $\kappa\rightarrow 0$ and using
the property that
\be
\wp(u)\sim u^{-2} \hskip 10pt \mbox{as}\hskip 10pt
u\rightarrow 0
\ee
gives
\be
z=-1/s
\ee
in this limit.

Since the monopole with this Nahm data is spherically
symmetric, we only need to constructing the Higgs
field along an axis. Setting $(x_1,x_2,x_3)=(0,0,r)$
the linear equation (\ref{lin}) becomes
\be
\frac{d{\bf v}}{ds}+
\left[\begin{array}{cccccc}
(r-1/s)&0&0&0&0&0\\
0&-(r-1/s)&-\sqrt{2}/s&0&0&0\\
0&-\sqrt{2}/s&r&0&0&0\\
0&0&0&-r&-\sqrt{2}/s&0\\
0&0&0&-\sqrt{2}/s&(r+1/s)&0\\
0&0&0&0&0&-(r+1/s)
\end{array}
\right]
{\bf v}=0.
\ee
Clearly the first and last components of ${\bf v}$
decouple and are elementary to solve. The remaining
four equations may be decoupled into two pairs, which
can then be converted into two second order equations
and solved by a Laplace transform. The full regular solution is
\be
{\bf v}=\left[\begin{array}{c}
c_1se^{-rs}\\
\\
c_2\sqrt{2}(\frac{\cosh(rs)}{rs}-\frac{\sinh(rs)}{r^2s^2})\\
\\
c_2(\frac{\sinh(rs)}{r^2s^2}-e^{-rs}(1+\frac{1}{rs}))\\
\\
c_3(-\frac{\sinh(rs)}{r^2s^2}-e^{rs}(1-\frac{1}{rs}))\\
\\
-c_3\sqrt{2}(\frac{\cosh(rs)}{rs}-\frac{\sinh(rs)}{r^2s^2})\\
\\
c_4se^{rs}
\end{array}
\right]
\ee
where $c_1,c_2,c_3,c_4$ are arbitrary constants.
If we select the orthonormal basis $\widehat {\bf v}_i$ 
where $\widehat {\bf v}_i$ has only $c_i$ non-zero, then 
the Higgs field will be diagonal.
Performing the required integrals (\ref{higgs}) gives
the result
\bea
(\Phi)_{11}&=&-i\ \frac{6r-3+e^{-8r}
(64r^3+48r^2+18r+3)}{2r(1-e^{-8r}(32r^2+8r+1))}\\
(\Phi)_{33}&=&i\ \frac{e^{8r}(-4r+3)-384r^3+64r^2-40r-6
+e^{-8r}(128r^3+128r^2+44r+3)}{e^{8r}(-4r+1)+128r^3
-8r-2+e^{-8r}(128r^3+64r^2+12r+1)}\nonumber\eea
with $(\Phi)_{22}$ and $(\Phi)_{44}$ obtained by the
replacement $r\mapsto -r$ in $(\Phi)_{11}$ and $(\Phi)_{33}$
respectively. It is a simple task to verify that indeed this
solution has the correct asymptotic behaviour ie.
\be
\Phi\rightarrow i\mbox{diag}(-3,1,1,1) \hskip 1cm
\mbox{as} \hskip 1cm r\rightarrow\infty.
\ee

To compute the Higgs field for non-zero values of
$\kappa$ is a much more difficult task, since all the
components of the vector ${\bf v}$ become coupled together.
Thus we turn to the numerical implementation of the ADHMN
construction. In Figure 1 we display the results in the form of
three dimensional plots of surfaces of constant energy density
for the values $\kappa/\kappa_0=-0.25,-0.10,0.00,+0.10,+0.25$.
As the  parameter $\kappa$ increases
from zero, the spherical monopole deforms into a tetrahedral
monopole. As $\kappa\rightarrow \pm\kappa_0/2$, 
the monopole approaches the embedded
 $SU(2)$ tetrahedral 3-monopole asymptotically. In fact,
even for the value  $\kappa=-\kappa_0/4$ (Figure 1.1) the
energy density looks very similar to that of the
$SU(2)$ tetrahedral 3-monopole \cite{HSa}. Note that
changing the sign of $\kappa$ gives a monopole corresponding
to the dual tetrahedron.

\begin{figure}[tb]
\begin{center}
\epsfig{file=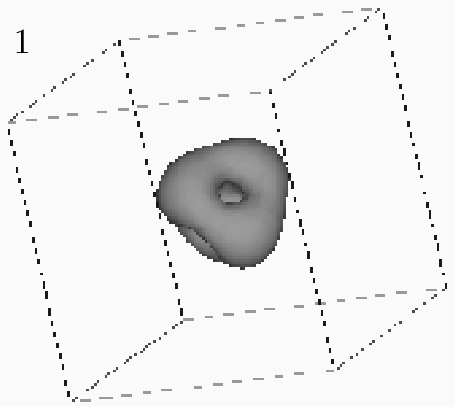,bbllx=55pt,bblly=674pt,bburx=208pt,bbury=787pt,
width=5cm}\qquad
\epsfig{file=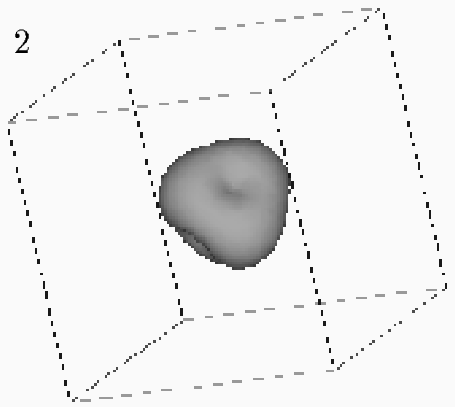,bbllx=55pt,bblly=674pt,bburx=208pt,bbury=787pt,
width=5cm}
\vskip .5cm
\epsfig{file=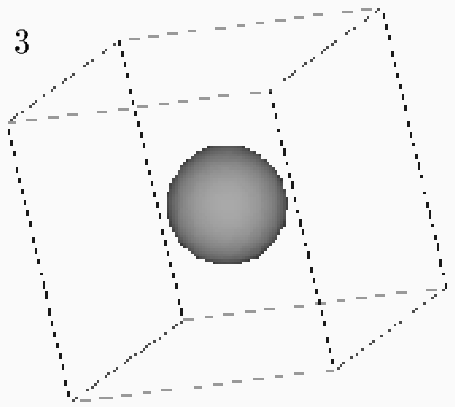,bbllx=55pt,bblly=674pt,bburx=208pt,bbury=787pt,
width=5cm}
\vskip .5cm
\epsfig{file=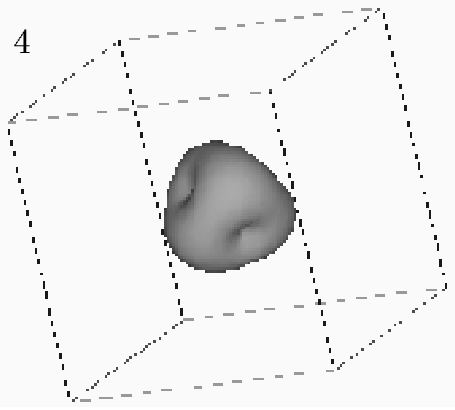,bbllx=55pt,bblly=674pt,bburx=208pt,bbury=787pt,
width=5cm}\qquad
\epsfig{file=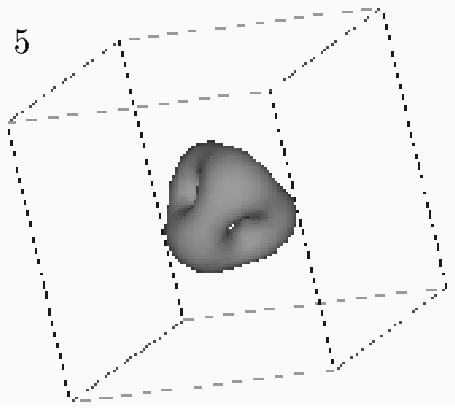,bbllx=55pt,bblly=674pt,bburx=208pt,bbury=787pt,
width=5cm}
\vskip  .5cm
\caption[Tetrahedral scattering of an $SU(4)$ 3-monopole]{Tetrahedral
  scattering of an $SU(4)$ 3-monopole.\label{tSU4fig}}
\end{center}
\end{figure}

In the moduli space approximation \cite{M} the dynamics
of $k$ monopoles is approximated
by geodesic motion on the $k$-monopole moduli space ${\cal M}_k$.
From the spectral curve approach it is clear that, having fixed
an orientation and centre of mass, we have constructed the unique
one-parameter family of tetrahedrally symmetric $SU(4)$
3-monopoles. Since the fixed point set of a group action
gives a totally geodesic submanifold, this one-parameter
family is a geodesic in ${\cal M}_3$. Hence, within the
moduli space approximation, this family of monopoles may
be interpreted as describing the low energy dynamics of
three deforming monopoles. Using this interpretation we
see from Figure 1 that during the course of the motion a
tetrahedron gets smoothed out into a sphere which then
deforms back into the dual tetrahedron. This gives an example of
dynamics in which the monopoles never become separated. 

It is known in the case of $SU(2)$ monopoles that there are
closed geodesics \cite{BM,B,HS5}. In the geodesics approximation such 
geodesics correspond to periodic monopole motions during which the
monopoles never separate. That is not the case here, here the motion
is not periodic and the geodesic is not closed; it runs between points
on the asymptotic boundary of the moduli space which do not correspond
to separated monopoles.

Obviously the method applied in this Section to 
tetrahedral $SU(4)$ 3-monopoles can easily be  carried
over to construct $SU(k+1)$ $k$-monopoles, given the
Nahm data for an $SU(2)$ $k$-monopole. In general, given
a $p$-dimensional family of $SU(2)$ monopoles there will
be a corresponding $(p+1)$-dimensional family of $SU(k+1)$
monopoles. Thus, for example, it is a simple task to construct the
Nahm data for the one-parameter family of octahedrally symmetric 
$SU(5)$ 4-monopoles which derive from the unique
octahedrally symmetric $SU(2)$ 4-monopole \cite{HMM}.
However, it is more interesting to consider
geodesic motion in the two-dimensional moduli space
of $SU(k+1)$ monopoles derived from a one-parameter
family of $SU(2)$ monopoles corresponding to a geodesic
in the $SU(2)$ moduli space. Physically, this will allow
us to examine how the dynamics of $SU(k+1)$ monopoles
compares with the dynamics of $SU(2)$ monopoles.
We shall do this in the following Section, for the
case of $SU(5)$  4-monopoles with tetrahedral symmetry.

\section{SU(5) monopoles with tetrahedral symmetry}\label{Smwts2}
\news
After fixing the orientation and centre of mass, there
is a one-parameter family of tetrahedrally symmetric
charge four $SU(2)$ monopoles \cite{HSa}. The associated Nahm
data takes the form
\be
T_i(s)=x(s)X_i+y(s)Y_i+z(s)Z_i \hskip 20pt i=1,2,3
\label{tnd}
\ee
where the tetrahedrally symmetric Nahm triplets are
\bea
(X_1+iX_2,X_3)&=&\left(
2\left[{\begin{array}{cccc}
0&0&0&0\\
-\sqrt{3}&0&0&0\\
0&-2&0&0\\
0&0&-\sqrt{3}&0 \end{array}}\right],
\left[{\begin{array}{cccc}
3i&0&0&0\\
0&i&0&0\\
0&0&-i&0\\
0&0&0&-3i \end{array}}\right]
\right)
\\
(Y_1+iY_2,Y_3)&=&
4
\left(
\left[{\begin{array}{cccc}
0&0&0&-5\\
\sqrt{3}&0&0&0\\
0&-3&0&0\\
0&0&\sqrt{3}&0 \end{array}}\right],
\left[{\begin{array}{cccc}
i&0&0&0\\
0&-3i&0&0\\
0&0&3i&0\\
0&0&0&-i \end{array}}\right]
\right)
\nonumber\\
(Z_1+iZ_2,Z_3)&=&\sqrt{3}\left(
2\left[{\begin{array}{cccc}
0&i&0&0\\
0&0&0&0\\
0&0&0&-i\\
0&0&0&0 \end{array}}\right],
\left[{\begin{array}{cccc}
0&0&1&0\\
0&0&0&1\\
-1&0&0&0\\
0&-1&0&0 \end{array}}\right]
\right)
\nonumber\eea
The reduced equations for the three real functions
$x,y,z$ can be solved to yield
\begin{eqnarray} x(s)&=&\frac{\kappa}{5}\left(-2\sqrt{\wp(\kappa
    s)}+\frac{1}{4}\frac{\wp^\prime(\kappa s)}{\wp(\kappa
    s)}\right)\label{xsoln}\\
y(s)&=&\frac{\kappa}{20}\left(\sqrt{\wp(\kappa
    s)}+\frac{1}{2}\frac{\wp^\prime
(\kappa s)}{\wp(\kappa s)}\right)\label{ysoln}\\
z(s)&=&\frac{a\kappa}{2\wp(\kappa s)}.
\label{zsoln}\end{eqnarray}
Here $\wp$ is the Weierstrass function satisfying
\be \wp^{\prime 2}=4\wp^3-4\wp+12a^2
\label{wfun}\ee
with prime denoting differentiation with respect to the argument.

The spectral curve for tetrahedrally symmetric 4-monopoles
has the form
\be
\eta^4+ i\alpha\eta\zeta(\zeta^4-1)+
\beta^2(\zeta^8+14\zeta^4+1)=0\label{sc}
\ee
where $\alpha$ and $\beta$ are real constants.
The relation between
these constants and those appearing in the above Nahm data
 is given by
\be
\alpha=36a\kappa^3, \hskip 1cm \beta^2=3\kappa^4.
\label{ab}
\ee

In the $SU(2)$ case, the requirement that the Nahm data
has a second pole at $s=2$ means that $\kappa$ must be
taken to be half the real period of the elliptic function
(\ref{wfun}). Thus, $\kappa$ is determined given the
parameter $a$, and we have the required one-parameter
family. Furthermore, $a$ is restricted to
lie in the interval $a\in(-a_c,a_c)$,
with $a_c=3^{-5/4}\sqrt{2}$. The elliptic function 
becomes rational at $a=\pm a_c$, with infinite real period
so that there is no second pole, and hence there is
no corresponding $SU(2)$ monopole.

Applying the boundary conditions for $SU(5)$ monopoles
is a different story: we now require no singularities
of the Nahm data for $s\in(0,5]$. If we consider
$a\in(-a_c,a_c)$ then the result is similar to that of
the previous Section. The range of $\kappa$
is now restricted, $\kappa\in(-\kappa_0,\kappa_0)$ where
$5\kappa_0$ is the real period of the elliptic
function (\ref{wfun}). Using the formula (\ref{ab})
this determines a domain in the $(\alpha,\beta)$ plane of
the spectral curve coefficients. For $a=\pm a_c$, there
is no second pole of the elliptic function, so the value
of $\kappa$ is unrestricted. This case corresponds to two
curves in the $(\alpha,\beta)$ plane, which pass through
the origin and continue off to infinity. 

To examine the case $\vert a\vert>a_c$ we need to
consider some properties of elliptic functions \cite{DV}.
For $\vert a\vert<a_c$ the
discriminant of the elliptic curve determined by (\ref{wfun})
is positive and the period lattice is rectangular.
The elliptic function has poles on the real axis, but no
zeros. However, for $\vert a\vert>a_c$ the character
of the elliptic function changes since the discriminant
is now negative. The period lattice is
rhombic and in addition to having poles on the
real axis, the elliptic function also has zeros on the
real axis. From equations (\ref{xsoln}-\ref{zsoln}) we
see that a zero of the elliptic function also corresponds
to singular Nahm data. Thus, in this case, there is 
a restriction on $\kappa$ given by 
$\kappa\in(-\widetilde\kappa_0,\widetilde\kappa_0)$ where
$5\widetilde\kappa_0$ is the smallest real root of
the elliptic function (\ref{wfun}), that is,
$\wp(5\widetilde\kappa_0)=0.$ This defines a second
domain in the $(\alpha,\beta)$ plane which matches smoothly
onto the first, with the joining boundary being the
curves determined by $a=\pm a_c$. 

We now have no restriction on the parameter
$a$, so we must also consider the limit $a\rightarrow\infty$.
In this limit it can be shown that 
$\widetilde\kappa_0\rightarrow 0$, but in such a way 
that the combination $a\widetilde\kappa_0^3$ is finite,
though it can be non-zero. In terms of the spectral
curve constants this limit corresponds to monopoles with
$\beta=0$, but $\alpha$ restricted only to lie in some
finite range. We refer to such monopoles as purely tetrahedral,
since the octahedral term in the spectral curve is absent.
This is an interesting result, since no such
purely tetrahedral charge four monopoles occur in the
 $SU(2)$ theory. Of course, given the existence of
purely tetrahedral monopoles it is simple
to study the reduced Nahm equations directly in
the case $\beta=0$ and obtain the same result as above
without the need for limit taking.

From the above analysis it is seen that in order to
compute the Nahm data and calculate the domain of
definition in the $(\alpha,\beta)$ plane, numerical algorithms
must be employed to compute not only elliptic functions
and their derivatives  but also their periods and 
elliptic logarithms. In the $SU(2)$ case this task was
much easier, since for a rectangular period lattice
the required computations can be performed using
Jacobi elliptic functions with real arguments. However,
in the rhombic case this is not true, and it is better
to work directly with the Weierstrass function.
Standard algorithms are used which are based upon the AGM method
and truncated series \cite{C}.

\begin{figure}[tb]
\begin{center}
\leavevmode
\epsfxsize=14cm \epsffile{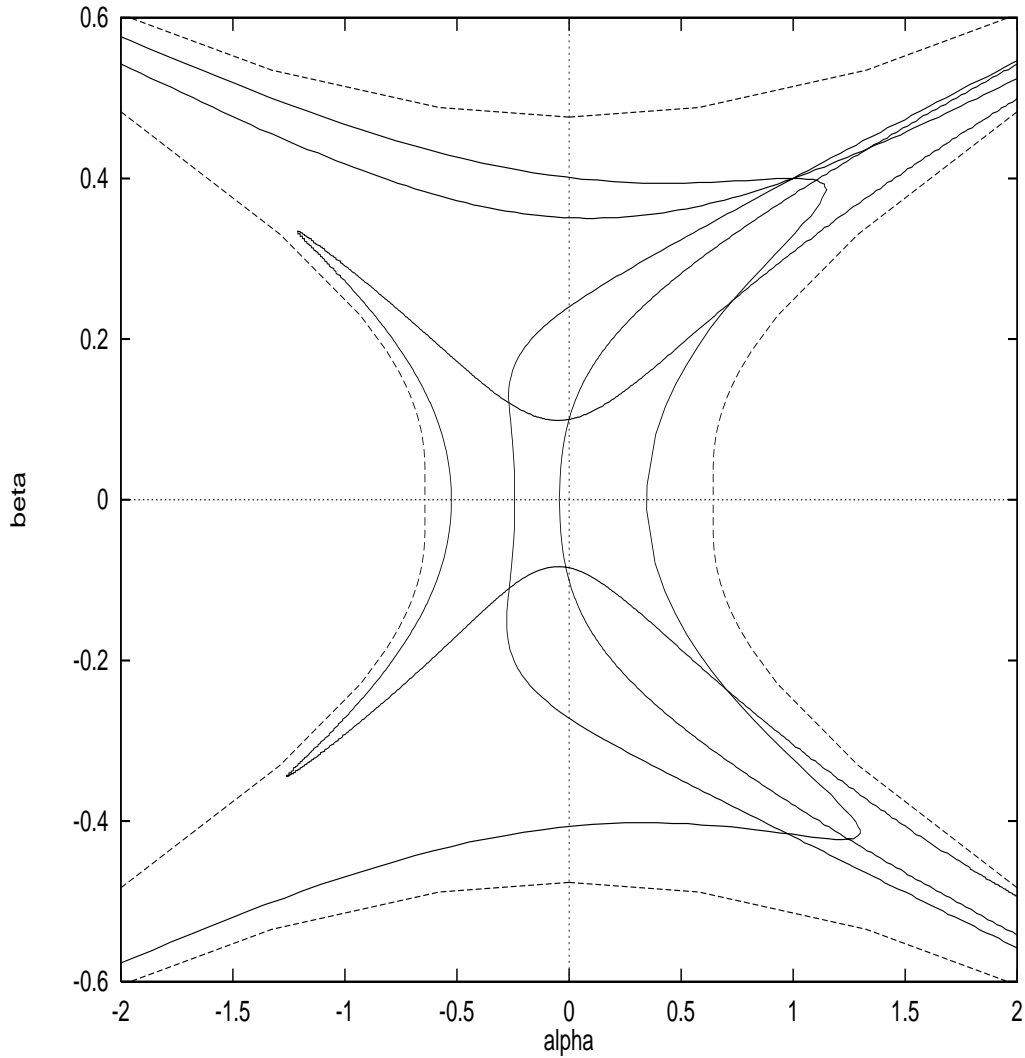}
\caption{Five geodesics for $SU(5)$ tetrahedral monopoles.}
\end{center}
\end{figure}

In Figure 2 we plot (dashed lines) the boundary of
the spectral curve coefficients in the $(\alpha,\beta)$ plane
for $\alpha\in[-2,2]$. Note that we allow $\beta$ to be
negative, even though the points $(\alpha,\beta)$ and
$(\alpha,-\beta)$ give the same spectral curve and
are hence gauge equivalent. The reason is that the
coefficient of the octahedral term in the spectral curve
is non-negative, so that $\beta^2$ is the correct 
parameterization, rather than, say, $\beta$.
This arises since a cube is inversion symmetric,
whereas the tetrahedron is not and
would lead to a change of sign in the spectral curve
coefficient. Thus the analogue of the $SU(4)$ tetrahedral
 geodesic of the last Section, where the tetrahedron deformed
to a sphere and out through the inverted tetrahedron, is the
$SU(5)$ octahedral geodesic along the line $\alpha=0$, where a
cube deforms into a sphere $(\alpha,\beta)=(0,0)$
and out through the inverted cube (which is gauge equivalent).

It should be stressed that our X-shaped representation
of the moduli space is not a reflection of  any metric 
properties of the moduli space.
As mentioned above, we know that the line $\alpha=0$ is 
a geodesic (by symmetry), but in order to determine
more general geodesics we must first compute the metric on
this two dimensional moduli space, and then solve the geodesic
equations of motion. We shall do this using numerical techniques,
though we must also make an assumption, as follows.

For $SU(2)$ monopoles it is known that the 
transformation between the  monopole 
moduli space metric and the metric on Nahm data is an 
isometry \cite{N}. However, for general $SU(N)$ gauge
groups it has not yet been proved that this transformation
is an isometry, although this is widely believed to be true.
There is also circumstantial evidence, for example,
assuming this result leads to monopole metrics which
reproduce conjectured metrics based upon asymptotic knowledge
 \cite{Mu}. To make progress we shall assume that this
transformation is an isometry, and compute the metric on
the Nahm data. This assumption was also made in previous
studies on the dynamics of $SU(3)$ monopoles \cite{DL}.

The scheme to compute the metric is similar to the
$SU(2)$ case \cite{S}, to which we refer the reader for
a more detailed discussion. Note that recently this $SU(2)$ 
metric has been computed exactly and in closed form \cite{BS}.
This was then used to demonstrate the excellent accuracy of the numerical
 algorithm \cite{S}.
It is likely that the method of \cite{BS} could also be used in this case 
to calculate the $SU(5)$ metric exactly, if required.

The tangent space is computed by solving the linearized
Nahm equation
\be
\frac{dV_i}{ds} =\epsilon_{ijk}[T_j,V_k]
\hskip 20pt i=1,2,3
\label{lne}
\ee
for the tangent vector $(V_1,V_2,V_3)$ corresponding
to the point with Nahm data $(T_1,T_2,T_3)$. Given
two tangent vectors $V_i,W_i$, the metric on Nahm data
is 
\be
<V_i,W_i>=-\int_0^5\sum_{i=1}^3\mbox{tr}(V_iW_i)\ ds.
\label{tvip}
\ee

From the tetrahedral symmetry of the Nahm data
it follows that the tangent
vectors are tetrahedrally symmetric so we may write
\be
V_i=q_1X_i+q_2Y_i+q_3Z_i \hskip 20pt i=1,2,3
\label{ttv}
\ee
where ${\bf q}=(q_1,q_2,q_3)^t$ is an analytic real 3-vector 
function of $s\in[0,5]$. In terms of 
${\bf q}$ equation (\ref{lne}) is
\be
\dot{\bf q}=M{\bf q} \hskip 20pt \mbox{where} \hskip 10pt
M=\left[{\begin{array}{ccc}
4x&-96y&-12z/5\\
-6y&-16y-6x&-6z/5\\
-4z&-32z&-4x-32y
\end{array}}\right].
\label{qode}
\ee
This ordinary differential equation has a
 regular-singular 
point at $s=0$. Analysis of the initial value problem
at $s=0$ reveals that there is a two-dimensional family 
of solutions
 which are normalizable for $s\in[0,5]$. They are given
by the two-parameter, ${\bf c}=(c_1,c_2)$, family of initial
conditions
\be
{\bf q}\sim(0,c_1 s^3,c_2 s^2)^t \hskip 10pt 
\mbox{as} \hskip 10pt s\sim 0.
\label{icl}
\ee

Using the asymptotic properties of the Weierstrass function
we find that the Nahm data has the behaviour
\be
y\sim \frac{\beta^2 s^3}{120}, \hskip 10pt
 z\sim \frac{\alpha s^2}{72} \hskip 10pt
\mbox{as} \hskip 10pt s\sim 0.
\ee
Hence to compute the tangent vector
$\frac{\partial}{\partial\alpha}$ dual to the coordinate
$\alpha$ requires the choice
${\bf c}=(0,1/72)$, whereas to compute the tangent
vector $\frac{\partial}{\partial\beta}$ dual to the coordinate
$\beta$ requires ${\bf c}=(\beta/60,0)$.
The metric can then be computed as
\bea
g_1&=&<\frac{\partial}{\partial\alpha},
\frac{\partial}{\partial\alpha}>\nonumber\\
g_2&=&<\frac{\partial}{\partial\beta},
\frac{\partial}{\partial\beta}>\\
g_3&=&<\frac{\partial}{\partial\alpha},
\frac{\partial}{\partial\beta}>
\eea
with corresponding Lagrangian
\be
{\cal L}=g_1(\frac{d\alpha}{dt})^2
+g_2(\frac{d\beta}{dt})^2
+2g_3(\frac{d\alpha}{dt})(\frac{d\beta}{dt}).
\label{lag}
\ee
The metric is computed numerically by solving
equation (\ref{qode}) using a fixed-step
fourth-order Runge-Kutta method, with the integrations
required in equation (\ref{tvip}) calculated via a
composite Simpsons rule. The geodesic equations which
follow from the Lagrangian (\ref{lag}) are solved
using a variable-step Runge-Kutta method, with the
derivatives of the metric approximated by finite differences.
The accuracy of our scheme was such that energy was
conserved to four significant figures for all computed
geodesic trajectories.

Note from the above that the metric components
$g_2$ and $g_3$ both vanish for $\beta=0$, which
is a reflection of our choice of $\beta^2$ as the
spectral curve coefficient and implies that all
geodesics which cross the $\alpha$-axis are
parallel to the $\beta$-axis at the point of crossing.
Thus from the numerical point of view we work with
the coordinate $\beta^2$ when computing geodesics,
since it is better behaved than $\beta$.

In Figure 2 we show five geodesics (solid lines),
three of which pass through the point 
$(\alpha,\beta)=(1.0,0.4)$ and the remaining two
pass through the point $(\alpha,\beta)=(0.0,0.1)$.
Many other geodesics were also computed, but the
qualitative features are captured by those shown.
Basically, the results show two kinds of scattering
that take place. The first kind is similar to the 
$SU(2)$ scattering and occurs when the geodesic does
not stray too far away from the $SU(2)$ embedding boundary.
The four monopoles approach from infinity on the vertices 
of a large
contracting tetrahedron, scatter through a cubic monopole, that is,
cross the $\beta$-axis, and emerge on the vertices
of an expanding tetrahedron dual to the incoming one.
We show one geodesic of this kind in the upper half
plane.
The second kind of scattering is more exotic and involves
a double scattering through a cubic monopole. 
The remaining four geodesics are all of this kind,
with three associated with monopoles which approach
from infinity with $\alpha$ positive and one with $\alpha$
negative. In each case the geodesic first crosses
the $\beta$-axis (a cubic scattering) and then crosses
the $\alpha$-axis, instantaneously forming a purely
tetrahedral monopole, after which it recrosses the
 $\beta$-axis (the second cubic scattering)
and goes off to infinity gauge equivalent to the
incoming configuration.

The two types of scattering described above were the only
ones found; no geodesics were found with, for example, no
cubic scatterings or more than two cubic scatterings.
In fact the results in this case are similar
in spirit to those seen in the study of $SU(3)$ 2-monopole
dynamics \cite{DL}, where it was found that up to two
$90^\circ$ scatterings could take place. It would therefore
seem that this phenomenon of multiple scatterings
is the generic situation for general $SU(N)$ monopoles
with minimal symmetry breaking.

\section*{Acknowledgements}

PMS acknowledges support from the Nuffield Foundation.  CJH
thanks the EPSRC and the British Council for financial support.


\newpage

\begin{thebibliography}{99}
\bibitem{BW} F.A. Bais and D. Wilkinson, Phys. Rev. D {\bf 19},
2410 (1979).
\bibitem{M} N.S. Manton, Phys. Lett. {\bf 110B}, 54 (1982).
\bibitem{W} E.J. Weinberg, Nucl. Phys. {\bf B167}, 500 (1980).
\bibitem{Wa} R.S. Ward, Commun. Math. Phys. {\bf 86}, 437 (1982).
\bibitem{D} A.S. Dancer, Commun. Math. Phys. {\bf 158}, 545 (1993).
\bibitem{DL} A.S. Dancer and R.A. Leese, Proc. R. Soc. London A
 {\bf 440}, 421 (1993).
\bibitem{Nahm} W. Nahm, \lq{\sl The construction of all self-dual
multimonopoles by the ADHM method}\rq, in Monopoles in quantum field
theory, eds. N.S. Craigie, P. Goddard and W. Nahm, (World Scientific,
Singapore, 1982).
\bibitem{HM} J. Hurtubise and M.K. Murray, Commun. Math. Phys. {\bf 122},
  35 (1989).
\bibitem{Na} H. Nakajima, \lq{\sl Monopoles and Nahm's
 equation}\rq, talk presented at the British Mathematical Colloquium,
UMIST, 1996.
\bibitem{HSa} C.J. Houghton and P.M. Sutcliffe,
  Commun. Math. Phys. {\bf 180}, 343 (1996).
\bibitem{HMM} N.J. Hitchin, N.S. Manton and M.K. Murray,
Nonlinearity {\bf 8}, 661 (1995).
\bibitem{HSb} C.J. Houghton and P.M. Sutcliffe, Nucl. Phys. {\bf
    B464}, 59 (1996).
\bibitem {BM} L. Bates and R. Montgomery, Commun. Math. Phys. 118, 635
  (1988). 
\bibitem{B} R. Bielawski, Nonlinearity 9, 1463 (1996).
\bibitem{HS5} C.J. Houghton and P.M. Sutcliffe, Nonlinearity 9, 1609 (1996).
\bibitem{DV} P. Du Val, \lq{\sl Elliptic functions and
elliptic curves}\rq, (Cambridge University Press, Cambridge 1973).
\bibitem{C} H. Cohen,  \lq{\sl A course in computational
algebraic number theory}\rq, (Springer-Verlag, Berlin, 1991).
\bibitem{N} H. Nakajima, \lq{\sl Monopoles and Nahm's
 equations}\rq, proceedings, Einstein metrics
and Yang-Mills connections, Sanda 1990, (Marcel Dekker, New York, 1993).
\bibitem{Mu} M.K. Murray, \lq{\sl A note on the (1,1,..,1)
monopole metric}\rq, hep-th/9605054.
\bibitem{S} P.M. Sutcliffe, Phys. Lett. {\bf 357B}, 335 (1995).
\bibitem{BS} H.W. Braden and P.M. Sutcliffe, Phys. Lett. {\bf B391}
  366, 1997 


\end{thebibliography}
\end{document}